\documentclass[twocolumn,floatfix,preprintnumbers,groupedaddress]{revtex4-1}
\pdfoutput=1

\usepackage{amsmath}
\usepackage{amssymb}
\usepackage{graphicx}
\usepackage{float}
\usepackage{braket}
\usepackage{hyperref}
\usepackage{pbox}
\usepackage{tabularx}

\usepackage[all]{hypcap}

\usepackage{feynmp}
\def\beq{\begin{equation}}
\def\eeq{\end{equation}}
\def\beeq{\begin{eqnarray}}
\def\eeeq{\end{eqnarray}}

\def\gp2{g^{\prime 2}}

\newcommand{\pt}{\partial}
\newcommand{\om}{\omega}

\newcommand{\bo}{\mathrm{O}}

\newcommand{\dif}{\mathrm{d}}

\newcommand{\del}{\delta}

\usepackage{color}

\begin{document}   
\title{Quantum Field Theory of Fluids} 
\date{\today}

\author{Ben Gripaios} 
\email{gripaios@hep.phy.cam.ac.uk}
\author{Dave Sutherland}
\email{dws28@cam.ac.uk}
\affiliation{Cavendish Laboratory, J.J. Thomson Ave, Cambridge, CB3
  0HE, UK}

\begin{abstract} 
The quantum theory of fields is largely
based on studying
  perturbations around non-interacting, or free, field theories,
  which correspond to a collection 
of quantum-mechanical harmonic oscillators. The quantum theory of an ordinary
fluid is `freer', in the sense that the non-interacting theory 
also contains an infinite
collection of  quantum-mechanical free particles, corresponding to
vortex modes.
By computing a variety of
correlation functions at tree- and loop-level, we give evidence that a
quantum perfect fluid can be
consistently formulated as a low-energy, effective field theory.
We speculate that the quantum behaviour is
  radically different to both classical fluids and quantum fields.
\end{abstract}   
\maketitle 





\section{Introduction \label{sec:intro}}
Fluids are ubiquitous in everyday life and were, arguably, the
prototypical example of a classical field theory in physics.
As such, it is natural
  to want to quantize them, as we have successfully done with
  many other classical fields. Since fluid
  behaviour is known to arise in systems with very different microscopic
  constituents, we expect that, at best, such a theory 
will take the form of a non-renormalizable, effective field
  theory (EFT), valid only at large enough distance and time scales,
  and the goal is to show that such a description exists.

 In trying to do
  so, one immediately encounters an obstruction in the form of
 fluid  vortices, which, classically, can have arbitrarily low energy,
  irrespective of their spatial extent \footnote{Those who read while in the bath will be able to verify this easily, by slowing
    stirring the water in circles of varying size.}. As we
  shall see below, this means that these excitations behave nothing
  like the infinite collection of harmonic oscillators that are the usual starting
  point for quantum field theory (QFT); instead they behave like a
  collection of quantum-mechanical free particles.

Landau, who was one of the first to attack the problem, tried to bypass the
obstruction by arguing 
  \cite{Landau1,Landau2} that the 
  vortex modes should be gapped in the quantum theory. In doing so, he
   stumbled not upon a quantum theory of fluids, but rather upon the theory of superfluids.

More recently, Endlich {\em et al.}\ \cite{Endlich:2010hf} conjectured that it is impossible
to quantize fluids. If true, this explains at a stroke why, in all known
 real-world examples, fluid behaviour does not persist to arbitrarily low
  temperatures ({\em e.g.}\ $\mathrm{H_2 O}$ freezes and $\mathrm{He}$ becomes
  superfluid): quantum effects must predominate eventually and so any classical fluid must change its phase before this happens.
The conjecture was supported by computations of $S$-matrix elements for
a putative quantum fluid, many of which turned out to diverge,
apparently making the `theory' useless \footnote{In fact, Endlich
  {\em et al.}\ focussed on the consequent
breakdown of unitarity;
 it seems to us that the divergence of tree-level $S$-matrix
  elements is a more fundamental problem {\em per se}.}. 

Here, we make a different conjecture, which is that quantum
  fluids are consistent, but that the peculiarities of  quantum
  mechanics make their phenomena completely different to those of
  classical fluids. 
If true, there might already exist real-world examples of quantum fluids,
without us even realizing it.
We support our conjecture by computing various
  correlation functions (`correlators') at tree- and loop-level and showing that
  they are well behaved. 

Our formulation of the problem, which we describe in \S \ref{sec:form},
largely follows that of \cite{Endlich:2010hf}, except that we work in
2+1-d spacetime, where we find a number of technical
simplifications. (There is no obstruction to carrying out the same
calculations for 3+1-d fluids, however, and we conjecture that
these are also consistent.) The key point of departure with \cite{Endlich:2010hf} is
  that we assert that, in a general physical theory, only quantities
  that are invariant under the symmetries of the theory are
  observable \footnote{Considering invariants
    was suggested in \cite{Endlich:2010hf}, but was not followed up}. This is a tautology, once we
  define the symmetries of a theory as those transformations that leave
  a system unchanged, and hence are unobservable. There are, of
  course, plenty of examples in physics where we can consistently compute non-invariants and use
  these as proxies for observables, but there are also plenty of
  examples where we cannot: gauge theories and 2-d sigma models are
  well-known examples. The $S$-matrix elements in these examples suffer from infra-red
  (IR) divergences that cancel when one computes correlators of
  invariants, {\em
    viz.}\ observables. 
Although we are unable to give a general
  proof, we will give multiple examples in \S \ref{sec:ir} where the
  same happens for fluids. 

Good IR behaviour alone does not suffice to establish consistency of the theory, however. Just like
in ordinary QFT, there are also ultraviolet (UV) divergences, 
coming from loop diagrams, and these must also be cancellable. Since the theory is
non-renormalizable, this requires, in general, an infinite tower of counterterms coming
from an expansion of the lagrangian in operators of increasing powers
of energy and momentum. This expansion will only `converge' in some
region of low energies and momenta, outside of which
predictivity is necessarily lost. To establish consistency, we must
show that such a region exists. Again, a general proof is
beyond us, but we do show, by a direct loop computation in a simple
example  in \S \ref{sec:uv}, that the necessary UV
cancellations occur, and that there exists a region of
energies and momenta where the expansion appears to be valid.
We speculate briefly on the implications in \S \ref{sec:disc}.
\section{Fluid parameterization \label{sec:form}}
We begin by discussing how to parameterize a fluid and its
dynamics. In the eulerian frame, a fluid is a
time-dependent map $\phi^i (x^j,t)$ from some space manifold $M$
(which we take to be $\mathbb{R}^2$) into itself.
We suppose that cavitation or interpenetration of the fluid costs finite
energy and may be ignored in our EFT description, such that $\phi$ is 1-to-1 and onto. 
Moreover, we assert that, by altering $\phi$ at short
distances, we can make it and its inverse smooth \footnote{If $M$
  is a
 torus, for example, this can be
  arranged by ensuring that the Fourier modes above the EFT cut-off
  fall off faster than any polynomial.},
such that $\phi$ is a diffeomorphism, and the configuration
space of the fluid is the diffeomorphism group $\mathrm{Diff} (M)$. We thus seek a parameterization of this group. 
$\mathrm{Diff}
(M)$ is infinite-dimensional and so is not a Lie group in the usual
sense;
the exponential map does not necessarily exist for non-compact $M$, and even for
compact $M$ it may not be locally-onto (indeed, $\mathrm{Diff}
(\mathbb{R})$ and $\mathrm{Diff} (S^1)$ are respective counterexamples
\cite{Khesin}). So, using
  the  na\"{\i}ve exponential map given in \cite{Endlich:2010hf} (which
  can be written as $\phi(x) = x + \pi + \frac{1}{2!} \pi \cdot \pt
  \pi + \frac{1}{3!} \pi \cdot \pt (\pi \cdot \pt \pi) + \dots$) is
  not necessarily adequate, even for small fluctuations. 
We therefore use the simple parameterization $\phi  = x +
  \pi$ (where $x$ is the identity map on $M$) and hope that all of the
  aforementioned demons are of measure zero in the path integral.

As for the dynamics, to have any chance of a quantum description
requires non-dissipative behaviour, so we assume the fluid to be
perfect \cite{[{For a recent attempt to incorporate viscous effects within a
  lagrangian formalism, see }]Endlich:2012vt}. The corresponding action has been
known for a long time \cite{herglotz}. It is most easily derived by requiring \cite{Endlich:2010hf}
that the theory be invariant under Poincar\'{e} transformations of $x$
\cite{[{For a formulation on a curved space, see }]Ballesteros:2012kv} and area-preserving diffeomorphisms of $\phi$. In 2+1-d, the lagrangian is
$\mathcal{L} = - w_0 f(\sqrt{B}),$
where $B = \mathrm{det} \ \pt_\mu \phi^i \pt^\mu \phi^j$, $f$ is
any function s.\ t.\ $f^\prime (1)=1$, and $w_0$ sets the overall dimension. Our
metric is mostly-plus and $\hbar$ and the speed of light are set to unity.
One may easily check that conservation of the energy-momentum
tensor, $T_{\mu \nu} = (\rho + p) u_\mu u_\nu + p \eta_{\mu \nu}$, (which for a fluid is equivalent to the Euler-Lagrange
equations \cite{soper1975}) holds with
$\rho = w_0 f$, 
$p =  w_0 ( \sqrt{B} f^\prime -f)$, and 
$u^\mu = \frac{1}{2\sqrt{B}}\epsilon^{\mu \alpha \beta} \epsilon_{ij}
\pt_\alpha \phi^i \pt_\beta \phi^j$.
In terms of $\phi^i = x^i + \pi^i$, we have
\begin{widetext}
\begin{multline}
\label{eq:2dlag}
\mathcal{L} = \frac{1}{2} (\dot\pi^2 - c^2 [\pt\pi]^2) - \frac{(3c^2 + f_3)}{6} [\pt\pi]^3 + \frac{c^2}{2}  [\pt\pi] [\pt\pi^2] + \frac{(c^2 +1)}{2}  [\pt\pi] \dot\pi^2 - \dot\pi \cdot \pt \pi \cdot \dot\pi 
-\frac{(f_4 + 3c^2 + 6f_3)}{24} [\pt\pi]^4 \\+ \frac{(c^2 + f_3)}{4} [\pt\pi]^2[\pt\pi^2] - \frac{c^2}{8} [\pt\pi^2]^2 
+ \frac{(1-c^2)}{8} \dot\pi^4 -c^2 [\pt\pi] \dot\pi \cdot \pt\pi \cdot \dot\pi - \frac{(1-3c^2-f_3)}{4} [\pt\pi]^2 \dot\pi^2 
+ \frac{(1-c^2)}{4} [\pt\pi^2] \dot\pi^2 + \frac{1}{2} \dot\pi \cdot
\pt\pi \cdot \pt\pi^T \cdot \dot\pi + \dots,
\end{multline}
\end{widetext}
where $f_n \equiv d^n f/d\sqrt{B}^n |_{B=1}$, $c \equiv \sqrt{f_2}$ is the
speed of sound, and $[\pt
\pi]$ is the trace of the matrix $\pt^i \pi^j$, {\em \&c}.
The obstruction to quantization is now evident: fields $\pi$ with
$[\pt \pi] = 0$,
corresponding to transverse fluctuations (or
infinitesimal vortices), have
no gradient energy, and correspond to quantum-mechanical free
particles, rather than harmonic oscillators. Thus, the energy
eigenvalues are continuous and there can be no particle intepretation
via Fock space. Even worse, the ground state is completely delocalized
in $\pi$, meaning that quantum fluctuations sample field
configurations where the interactions are arbitrarily large. It thus
appears that perturbation theory is hopeless! From the path-integral
point of view, these difficulties translate into the statement that
the spacetime propagator for transverse modes is ill-defined, since it
contains the Fourier transform $\int
d \om e^{i \om t} /\om^2$, which diverges in the IR.
\section{Infra-red behaviour \label{sec:ir}}
Just as for gauge theories and 2-d sigma models \cite{Coleman:1973ci,Jevicki:1977zn,Elitzur:1978ww,McKane:1979cm,David:1980gi,David:1980rr}, the
IR divergences cancel when we restrict to correlators of invariants
under $\mathrm{SDiff} (M)$, such as $\rho, p$, and
$u^i$\footnote{These quantities are not all
  Poincar\'{e} invariant, so they are still really only proxies for
  observables.}. We can check the cancellation
order-by-order in $1/w_0$ (which is equivalent to the usual $\hbar$
expansion of QFT) or indeed in any other parameter. 

For
the 2-point correlators at $O(w_0^{-1})$, the observables can be
expressed in terms of
$[\pt \pi]$ and $\dot \pi$, whose correlators are
\begin{align}
\label{eq:2pt}
\langle [\pt \pi][\pt \pi] \rangle &= 
\frac{ik^2}{\om^2 -c^2 k^2}, \nonumber \\
\langle \dot{\pi}^i [\pt \pi] \rangle &= \frac{i\om
  k^i}{\om^2 -c^2 k^2},
\nonumber \\
\langle \dot{\pi}^i \dot{\pi}^j \rangle &= 
i\delta^{ij} + \frac{ic^2 k^i k^j}{\om^2 -c^2 k^2}.
\end{align}
The only poles are at $\om = c k$ 
and the disappearance of poles at $\om = 0$ implies that the spacetime
Fourier transforms are well-defined. 

To check for cancellations of IR divergences at higher order in
$w_0^{-1}$, it is convenient to consider the invariants
\begin{align}
\label{eq:quadobs}
 \sqrt{B}u^0 -1 &=  [\pt \pi] + \frac{1}{2} ([\pt\pi]^2 - [\pt\pi^2]), \nonumber \\
 \sqrt{B}u^i &= \dot \pi^i + [\pt\pi] \dot \pi^i - \dot \pi^j \pt_j \pi^i,
\end{align}
since (in $2+1$-d) they contain terms of
at most quadratic order in $\pi$. 
Consider, for example, the 3-point
correlator $\langle \sqrt{B} u^i (x_1,t_1) \sqrt{B} u^j(x_2,t_2) (\sqrt{B} u^0(0,0)-1)\rangle$ at $O(w_0^{-2})$, connected with respect to the three observables. The four contributing diagrams and their divergent pieces are:
\begin{widetext}
\begin{tabular}{m{45mm} m{2cm} m{5mm} m{2cm} m{5mm} m{2cm} m{5mm} m{2cm}}
$\langle \sqrt{B} u^i \sqrt{B} u^j (\sqrt{B} u^0-1)\rangle = $&
\begin{fmffile}{ij0_feynmp}
\begin{fmfgraph*}(15,15)
\fmfleft{3}
\fmfright{2,1}
\fmf{plain}{3,v1}
\fmf{plain}{2,v1,1}
\fmflabel{}{1}
\end{fmfgraph*}
\end{fmffile}&
\hspace{2mm}$+$\hspace{2mm}&
\begin{fmffile}{ij02_feynmp}
\begin{fmfgraph*}(15,15)
\fmfleft{3}
\fmfright{2,1}
\fmf{plain}{3,1}
\fmf{plain}{1,2}
\fmflabel{}{1}
\end{fmfgraph*}
\end{fmffile}&
\hspace{2mm}$+$\hspace{2mm}&
\begin{fmffile}{ij03_feynmp}
\begin{fmfgraph*}(15,15)
\fmfleft{3}
\fmfright{2,1}
\fmf{plain}{3,2}
\fmf{plain}{2,1}
\fmflabel{}{1}
\end{fmfgraph*}
\end{fmffile}&
\hspace{2mm}$+$\hspace{2mm}&
\begin{fmffile}{ij04_feynmp}
\begin{fmfgraph*}(15,15)
\fmfleft{3}
\fmfright{2,1}
\fmf{plain}{3,2}
\fmf{plain}{3,1}
\fmflabel{}{1}
\end{fmfgraph*}
\end{fmffile}
\end{tabular}

\begin{align*}
 &= \frac{1}{\om^2_3 - c^2 k^2_3} \Bigg( \frac{c^2 k_3^2 - 2 \om_1^2}{2\om_1\om_2}  (k_1T_2)^j (k_2T_1)^i + \om_1 \frac{(k_3 k_2)}{\om_2} (T_1 T_2)^{ij} + \frac{(k_1T_2)^j}{\om^2_1 - c^2 k^2_1} \frac{k^i_1}{k^2_1} \frac{\om_1}{\om_2} \left( (c^2 k^2_3-\om_1^2)  (k_2 k_1) - (c^2 k_1^2-\om_1^2) (k_2 k_3) \right) \\
& + [\{1,i\}\leftrightarrow \{2,j\}] \Bigg) + \frac{\om_3}{\om_2} \frac{1}{\om^2_3 - c^2 k^2_3} (k_2 k_3) (T_2)^{ij} + \frac{\om_3}{\om_1} \frac{1}{\om^2_3 - c^2 k^2_3} (k_1 k_3) (T_1)^{ij} + \left( \frac{1}{\om_1\om_2} (k_1T_2)^j (k_2T_1)^i + \frac{\om_1}{\om_2} \frac{k_1^i}{k_1^2} \frac{(k_2 k_1) (k_1 T_2)^j}{\om_1^2 - c^2 k_1^2} \right),
\end{align*}
\end{widetext}
where $(k_a,\om_a), a \in \{1,2\}$ are the Fourier conjugates of
$(x_a,t_a)$, $\om_3 = \om_1 + \om_2$, {\em \&c}. We define the
transverse projector by $T_a^{ij} \equiv \del^{ij} - \frac{k_a^i
  k_a^j}{k_a^2}$. Groups of $k$s or $T$s in brackets have their
indices contracted. It is clear that, by expansion about small
$\om_2$, $\frac{1}{\om^2_3 - c^2 k^2_3} = \frac{1}{\om^2_1 - c^2
  k^2_3} + \bo(\om_2)$ and the above poles at $\om_2=0$ cancel. By
symmetry, the
same is true for $\om_1$.

One may similarly show that divergences cancel in all 3-point
correlators of the observables in (\ref{eq:quadobs}). We have also
checked several 4-point tree-level correlators.
\section{Ultra-Violet behaviour \label{sec:uv}}
We now turn to loop diagrams. Consider, for example, the 2-point
function of $\sqrt{B} u^0 - 1$ at $\bo(w_0^{-2})$. The diagrams, shown in Fig.~\ref{fig:loopdiags}, feature both IR and UV divergences, which
we regularize by computing the integrals in $D=1+2\epsilon$ time- and
$d=2+2\epsilon$ space-dimensions. We wish to show that the UV
divergences can be absorbed in higher order counterterms and that
the expansion in energy and momenta is valid in some non-vanishing
region.

It is here that the advantage of working in $2+1$-d becomes clear: 
If the theory is to be consistent, the sum of the individually divergent diagrams
in Fig.~\ref{fig:loopdiags} must be finite as $\epsilon \rightarrow
0$, because there can be no counterterms! This follows from simple dimensional analysis:
the Feynman rules that follow from (\ref{eq:2dlag}) imply that the 1-loop diagrams
must contain 3 more powers of energy or momentum than the tree-level
diagrams. Now, since the correlator can only be a function of $K^2$
(where $ic K\equiv \om$)
and $k^2$ (by time-reversal and rotation invariance, respectively),
the 1-loop contribution necessarily contains radicals of $K^2$
and $k^2$. But higher order
counterterms can only yield tree-level contributions that are rational
functions of $K^2$
and $k^2$ and so cannot absorb divergences in the 1-loop
contribution. 

\begin{table*}

\begin{tabular}{c c c}
 $\int \frac{\dif^d p \dif^D P}{(4 \pi)^\frac{d+D}{2}} \frac{1}{P^2 + p^2} \frac{1}{(P+K)^2 + (p+k)^2}  \frac{1}{p^2} \frac{1}{(p+k)^2}$ & =&$\frac{1}{8 \pi  \epsilon k}+\frac{\alpha}{2 \pi  k}$ \\ 
 $\int \frac{\dif^d p \dif^D P}{(4 \pi)^\frac{d+D}{2}} \frac{1}{(P+K)^2 + (p+k)^2}  \frac{1}{p^2}$ & =&$\frac{1}{8 \sqrt{K^2+k^2}}$ \\ 
 $\int \frac{\dif^d p \dif^D P}{(4 \pi)^\frac{d+D}{2}} \frac{1}{P^2} \frac{1}{(P+K)^2} \frac{1}{p^2} \frac{1}{(p+k)^2}$ & =&$-\frac{1}{K^3 k^2}$ \\ 
 $\int \frac{\dif^d p \dif^D P}{(4 \pi)^\frac{d+D}{2}} \frac{1}{P^2 + p^2} \frac{1}{(P+K)^2}$ &=& $-\frac{3 \epsilon}{4 K}$ \\ 
 $\int \frac{\dif^d p \dif^D P}{(4 \pi)^\frac{d+D}{2}} \frac{1}{P^2 + p^2} \frac{1}{(p+k)^2}$ &=& $\frac{1}{8 \pi  \epsilon k}+\frac{\alpha}{2 \pi  k}$ \\ 
 $\int \frac{\dif^d p \dif^D P}{(4 \pi)^\frac{d+D}{2}} \frac{1}{P^2 + p^2} \frac{1}{(P+K)^2 + (p+k)^2}  \frac{1}{p^2}$ & =&$\frac{K^2-k^2}{8 \pi  \epsilon k \left(K^2+k^2\right)^2}+\frac{k}{2 \pi  \left(K^2+k^2\right)^2}+\frac{\alpha \left(K^2-k^2\right)}{2 \pi  k \left(K^2+k^2\right)^2}-\frac{2 K \tan ^{-1}\left(\frac{K}{k}\right)}{\pi  \left(K^2+k^2\right)^2}$ \\ 
 $\int \frac{\dif^d p \dif^D P}{(4 \pi)^\frac{d+D}{2}} \frac{1}{P^2 + p^2} \frac{1}{(P+K)^2 + (p+k)^2}  \frac{1}{p^2} \frac{1}{(p+k)^2}$ & =&$\frac{K^2-k^2}{4 \pi  \epsilon k^3 \left(K^2+k^2\right)^2}+\frac{2 \tan ^{-1}\left(\frac{k}{K}\right)}{\pi  K^3 k^2}+\frac{\alpha \left(K^2-k^2\right)}{\pi  k^3 \left(K^2+k^2\right)^2}+\frac{4 \left(2 K^2+k^2\right) \tan ^{-1}\left(\frac{K}{k}\right)}{\pi  K^3 \left(K^2+k^2\right)^2}-\frac{1}{K^3 k^2} -\frac{K^5+2 K^3 k^2+2 K k^4}{\pi  K^3 k^3 \left(K^2+k^2\right)^2}$ \\ 
 $\int \frac{\dif^d p \dif^D P}{(4 \pi)^\frac{d+D}{2}} \frac{1}{P^2 + p^2} \frac{1}{(P+K)^2} \frac{1}{p^2} \frac{1}{(p+k)^2}$ & =&$\frac{K^2-k^2}{8 \pi  \epsilon k^3 \left(K^2+k^2\right)^2}+\frac{\tan ^{-1}\left(\frac{k}{K}\right)}{\pi  K^3 k^2}+\frac{\alpha \left(K^2-k^2\right)}{2 \pi  k^3 \left(K^2+k^2\right)^2}+\frac{2 \left(2 K^2+k^2\right) \tan ^{-1}\left(\frac{K}{k}\right)}{\pi  K^3 \left(K^2+k^2\right)^2}-\frac{1}{2 K^3 k^2} -\frac{K^5+2 K^3 k^2+2 K k^4}{2 \pi  K^3 k^3 \left(K^2+k^2\right)^2}$ \\ 
 $\int \frac{\dif^d p \dif^D P}{(4 \pi)^\frac{d+D}{2}} \frac{1}{P^2 + p^2} \frac{1}{(P+K)^2} \frac{1}{(p+k)^2}$ & =&$\frac{K^2-k^2}{8 \pi  \epsilon k \left(K^2+k^2\right)^2}+\frac{k}{2 \pi  \left(K^2+k^2\right)^2}+\frac{\alpha \left(K^2-k^2\right)}{2 \pi  k \left(K^2+k^2\right)^2}-\frac{2 K \tan ^{-1}\left(\frac{K}{k}\right)}{\pi  \left(K^2+k^2\right)^2}$
\end{tabular}
\caption{Master integrals for the 1-loop, 2-point correlator with external momentum
  $k$ and euclidean energy $K$, dimensionally regularized with $d =2 +
  2\epsilon$, $D = 1+2\epsilon$, to
  $\bo(\epsilon^0)$; $\alpha(k^2) = \frac{1}{2} \log \left( \frac{2 e^{\gamma_E} k^2}{\pi} \right)$. The 4th integral appears with a $\frac{1}{\epsilon}$ coefficient in the correlator, and is expanded to $\bo(\epsilon^1)$.\label{tab:master}}
\end{table*}

\begin{figure}
\vspace{-10mm}
\begin{fmffile}{loop_feynmp}
\begin{fmfgraph*}(25,25)
\fmfleft{l1}
\fmfright{r1}
\fmf{plain}{l1,v1}
\fmf{plain}{v2,r1}
\fmf{phantom}{v1,v2}
\fmffreeze
\fmf{plain,left=1}{v1,v2}
\fmf{plain,right=1}{v1,v2}
\end{fmfgraph*}
\end{fmffile}
\begin{fmffile}{loop2_feynmp}
\begin{fmfgraph*}(25,25)
\fmfleftn{l}{21}
\fmfrightn{r}{21}
\fmftop{t1}
\fmfbottom{b1}
\fmf{phantom,tension=0.1}{t1,vdummy1}
\fmf{phantom,tension=0.1}{b1,vdummy2}
\fmf{plain,tension=0.5}{l10,vdummy2}
\fmf{plain,left=0.2}{vdummy1,v1,vdummy2}
\fmf{plain,tension=0.5}{vdummy1,l12}
\fmf{plain,tension=0.8}{v1,r11}
\end{fmfgraph*}
\end{fmffile}

\vspace{-10mm}
\begin{fmffile}{loop3_feynmp}
\begin{fmfgraph*}(25,25)
\fmfleftn{l}{21}
\fmfrightn{r}{21}
\fmftop{t1}
\fmfbottom{b1}
\fmf{phantom,tension=0.1}{t1,vdummy1}
\fmf{phantom,tension=0.1}{b1,vdummy2}
\fmf{plain,tension=0.5}{r10,vdummy2}
\fmf{plain,right=0.2}{vdummy1,v1,vdummy2}
\fmf{plain,tension=0.5}{vdummy1,r12}
\fmf{plain,tension=0.8}{v1,l11}
\end{fmfgraph*}
\end{fmffile}
\begin{fmffile}{loop4_feynmp}
\begin{fmfgraph*}(25,25)
\fmfleftn{l}{21}
\fmfrightn{r}{21}
\fmf{plain}{r10,l10}
\fmf{plain}{r12,l12}
\end{fmfgraph*}
\end{fmffile}
\vspace{-10mm}
\caption{The $\bo(w_0^{-2})$ diagrams for the correlator $\langle (\sqrt{B} u^0 - 1) (\sqrt{B} u^0 - 1) \rangle$.\label{fig:loopdiags}}
\end{figure}

To do the computation, we use integration-by-parts identities obtained using $\mathtt{AIR}$ \cite{air} to reduce the various loop integrals to
a set of 9 master integrals,
listed in Table~\ref{tab:master}.
All but the last 2 of these can be evaluated directly, in
terms of Gamma or Hypergeometric functions. For the remaining 2, we
proceed by deriving a first-order ODE for each
integral's dependence on $K^2$ and solving order-by-order in $\epsilon$. All the integrals were checked numerically in
dimensions where they are finite. Substituting in the loop amplitude
using $\mathtt{FORM}$ \cite{form},
we obtain
\begin{multline*}
\frac{9Kk^6(1+c^4)}{64(K^2+k^2)^2} 
-\frac{k^4}{1024 c^4 (K^2 +k^2)^\frac{5}{2}}\\
\times \Big[ c^4(1-c^2)^2 (19k^4 - 4K^2 k^2 + K^4)\\
-2f_3c^2 (1+c^2) k^2 (5k^2 +14K^2)
+ f_3^2(3k^4 + 8K^2 k^2 +8 K^4)
\Big], \nonumber
\end{multline*} 
which is indeed finite, as consistency demands. Moreover, there are no
poles at $K=0$ and the Fourier transform is well defined.

\begin{figure}
 \includegraphics{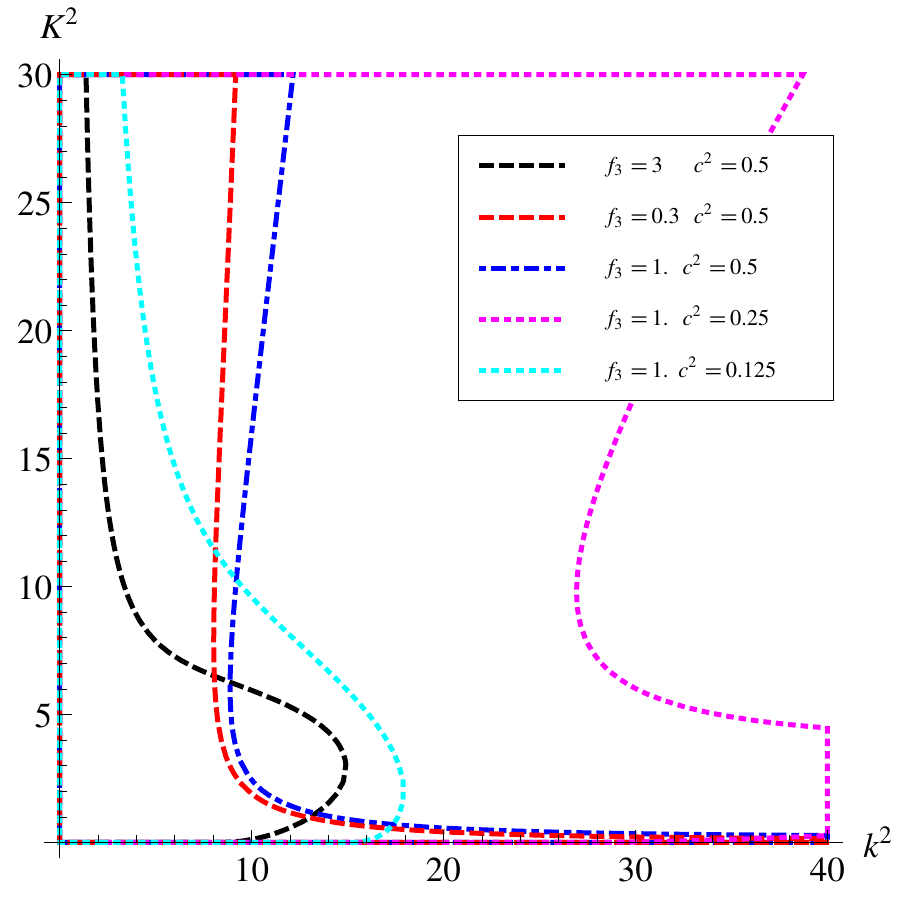}
\caption{Contours of equal 1-loop and tree-level absolute contributions to
  the momentum-space 2-point correlator $\langle (\sqrt{B} u^0 - 1)
  (\sqrt{B} u^0 - 1) \rangle$, for various $\bo (1)$ values of $c$ and $f_3$.\label{fig:valid}}
\end{figure}

Finally, we estimate the region of validity
of the EFT
expansion in energy-momentum, by comparing the absolute values of the tree-level and
1-loop results. Our estimate depends, of course, on the values of the
$O(1)$ coefficients $c^2$ and $f_3$, and we present results for
typical values (in units of the overall scale $w_0$) in Fig.~\ref{fig:valid}. It should be borne in
mind that this really constitutes only a rough upper bound on the
region of validity; in particular, we expect that comparison of other
diagrams will indicate that the EFT is not valid at arbitrarily
large energy, for small enough momentum (and {\em vice versa}), as the
Figure suggests.

\section{Discussion \label{sec:disc}}
Our results are a strong hint that there exists a consistent quantum
theory of fluids. If so, it is of great interest to explore the
physical predictions of the theory, and to see whether they are
realized in real-world systems. We can already draw some inferences
from the results derived here. The first of these is that Lorentz
invariance is non-linearly realized in the quantum vacuum, just as it
is in a classical fluid. This follows immediately from the occurrence
of poles at $\om = ck$ in the 2-point correlators
(\ref{eq:2pt}). Furthermore, the linearly realized symmetries appear
to be the same in the quantum theory as in the classical theory, {\em
  viz.}\ the diagonal euclidean subgroup of Poincar\'{e}$\times \mathrm{SDiff}$. The second is that vortex
modes apparently do not propagate, in the sense that they do not
appear as poles in correlators of observables. In hindsight this is no
surprise, since propagating vortices would imply IR divergences. We
stress, though, that the absence of vortex modes does not mean that
our fluid EFT is nothing but a complicated reformulation of a
superfluid. Indeed, it is already known that a superfluid 
and an ordinary fluid are inequivalent at $\hbar = 0$ (although they
are equivalent if there is no vorticity) \cite{Dubovsky:2005xd},
and it follows by continuity that fluids
and superfluids must be inequivalent in general at $\hbar \neq 0$.
It is tempting to conjecture, however, that both the conservation of vorticity
and the
equivalence between the zero-vorticity fluid and the superfluid are
preserved at the quantum level; if so, we must look to quantum fluids
with non-vanishing vorticity in order to see a departure from
superfluid behaviour. One possible arena would be the study of the
quanta corresponding to Kelvin waves \cite{Thomson}, {\em viz.} low-energy
perturbations of vortex lines \cite{[{For a recent lagrangian
  derivation of these, see }]Endlich:2013dma}, for which
`Thomsons' is the obvious moniker. More generally, it would be
of interest to explore the quantum version of any of the myriad
phenomena of classical fluids: surface waves, turbulence, shocks, {\em
  \&c}.

Where can we hope to observe such phenomena? 
Classical fluid behaviour is typically observed in underlying
systems that are in local thermodynamic equilibrium at finite
temperature. To see quantum behaviour in such a system, we would need
to somehow ensure that thermal fluctuations are negligible in the
long-distance fluid modes, which are what we quantize here. 
Alternatively, perhaps the correspondence of the theory with a fluid at
the classical level is a red herring. We have given evidence that
there exists an EFT, based on simple field content
and symmetries, with behaviour that is qualitatively novel. That is interesting enough in itself, and leads us to
hope that Nature may choose to make use of it {\em somewhere}. 

\section*{Acknowledgements}
BG acknowledges
the support of STFC, the
IPPP, and King's College,
Cambridge and thanks N.~Arkani-Hamed, B.~Bellazzini,  J.~Cardy, S.~Endlich, A.~Mitov, C.~Mouhot,
O.~Randall-Williams, R.~Rattazzi, and D.~Skinner for discussions. DS
acknowledges the support of STFC and Emmanuel College, Cambridge, and thanks T.~Gillam
and the authors of $\mathtt{xAct}$ \cite{xact} for computing help.
\bibliography{fluids}

\begin{thebibliography}{25}%
\makeatletter
\providecommand \@ifxundefined [1]{%
 \@ifx{#1\undefined}
}%
\providecommand \@ifnum [1]{%
 \ifnum #1\expandafter \@firstoftwo
 \else \expandafter \@secondoftwo
 \fi
}%
\providecommand \@ifx [1]{%
 \ifx #1\expandafter \@firstoftwo
 \else \expandafter \@secondoftwo
 \fi
}%
\providecommand \natexlab [1]{#1}%
\providecommand \enquote  [1]{``#1''}%
\providecommand \bibnamefont  [1]{#1}%
\providecommand \bibfnamefont [1]{#1}%
\providecommand \citenamefont [1]{#1}%
\providecommand \href@noop [0]{\@secondoftwo}%
\providecommand \href [0]{\begingroup \@sanitize@url \@href}%
\providecommand \@href[1]{\@@startlink{#1}\@@href}%
\providecommand \@@href[1]{\endgroup#1\@@endlink}%
\providecommand \@sanitize@url [0]{\catcode `\\12\catcode `\$12\catcode
  `\&12\catcode `\#12\catcode `\^12\catcode `\_12\catcode `\%12\relax}%
\providecommand \@@startlink[1]{}%
\providecommand \@@endlink[0]{}%
\providecommand \url  [0]{\begingroup\@sanitize@url \@url }%
\providecommand \@url [1]{\endgroup\@href {#1}{\urlprefix }}%
\providecommand \urlprefix  [0]{URL }%
\providecommand \Eprint [0]{\href }%
\providecommand \doibase [0]{http://dx.doi.org/}%
\providecommand \selectlanguage [0]{\@gobble}%
\providecommand \bibinfo  [0]{\@secondoftwo}%
\providecommand \bibfield  [0]{\@secondoftwo}%
\providecommand \translation [1]{[#1]}%
\providecommand \BibitemOpen [0]{}%
\providecommand \bibitemStop [0]{}%
\providecommand \bibitemNoStop [0]{.\EOS\space}%
\providecommand \EOS [0]{\spacefactor3000\relax}%
\providecommand \BibitemShut  [1]{\csname bibitem#1\endcsname}%
\let\auto@bib@innerbib\@empty
\bibitem [{Note1()}]{Note1}%
  \BibitemOpen
  \bibinfo {note} {Those who read while in the bath will be able to verify this
  easily, by slowing stirring the water in circles of varying
  size.}\BibitemShut {Stop}%
\bibitem [{\citenamefont {Landau}(1941{\natexlab{a}})}]{Landau1}%
  \BibitemOpen
  \bibfield  {author} {\bibinfo {author} {\bibfnamefont {L.~D.}\ \bibnamefont
  {Landau}},\ }\href {\doibase 10.1103/PhysRev.60.356} {\bibfield  {journal}
  {\bibinfo  {journal} {Phys. Rev.}\ }\textbf {\bibinfo {volume} {60}},\
  \bibinfo {pages} {356} (\bibinfo {year} {1941}{\natexlab{a}})}\BibitemShut
  {NoStop}%
\bibitem [{\citenamefont {Landau}(1941{\natexlab{b}})}]{Landau2}%
  \BibitemOpen
  \bibfield  {author} {\bibinfo {author} {\bibfnamefont {L.~D.}\ \bibnamefont
  {Landau}},\ }\href@noop {} {\bibfield  {journal} {\bibinfo  {journal} {J.
  Phys. USSR}\ }\textbf {\bibinfo {volume} {5}},\ \bibinfo {pages} {71}
  (\bibinfo {year} {1941}{\natexlab{b}})}\BibitemShut {NoStop}%
\bibitem [{\citenamefont {Endlich}\ \emph {et~al.}(2011)\citenamefont
  {Endlich}, \citenamefont {Nicolis}, \citenamefont {Rattazzi},\ and\
  \citenamefont {Wang}}]{Endlich:2010hf}%
  \BibitemOpen
  \bibfield  {author} {\bibinfo {author} {\bibfnamefont {S.}~\bibnamefont
  {Endlich}}, \bibinfo {author} {\bibfnamefont {A.}~\bibnamefont {Nicolis}},
  \bibinfo {author} {\bibfnamefont {R.}~\bibnamefont {Rattazzi}}, \ and\
  \bibinfo {author} {\bibfnamefont {J.}~\bibnamefont {Wang}},\ }\href {\doibase
  10.1007/JHEP04(2011)102} {\bibfield  {journal} {\bibinfo  {journal} {JHEP}\
  }\textbf {\bibinfo {volume} {1104}},\ \bibinfo {pages} {102} (\bibinfo {year}
  {2011})},\ \Eprint {http://arxiv.org/abs/1011.6396} {arXiv:1011.6396
  [hep-th]} \BibitemShut {NoStop}%
\bibitem [{Note2()}]{Note2}%
  \BibitemOpen
  \bibinfo {note} {In fact, Endlich {\protect \em et al.}\ focussed on the
  consequent breakdown of unitarity; it seems to us that the divergence of
  tree-level $S$-matrix elements is a more fundamental problem {\protect \em
  per se}.}\BibitemShut {Stop}%
\bibitem [{Note3()}]{Note3}%
  \BibitemOpen
  \bibinfo {note} {Considering invariants was suggested in \cite
  {Endlich:2010hf}, but was not followed up}\BibitemShut {NoStop}%
\bibitem [{Note4()}]{Note4}%
  \BibitemOpen
  \bibinfo {note} {If $M$ is a torus, for example, this can be arranged by
  ensuring that the Fourier modes above the EFT cut-off fall off faster than
  any polynomial.}\BibitemShut {Stop}%
\bibitem [{\citenamefont {Khesin}\ and\ \citenamefont {Wendt}(2009)}]{Khesin}%
  \BibitemOpen
  \bibfield  {author} {\bibinfo {author} {\bibfnamefont {B.}~\bibnamefont
  {Khesin}}\ and\ \bibinfo {author} {\bibfnamefont {R.}~\bibnamefont {Wendt}},\
  }\href@noop {} {\emph {\bibinfo {title} {{The Geometry of
  Infinite-Dimensional Groups}}}},\ \bibinfo {series} {{A Series of Modern
  Surveys in Mathematics}}, Vol.~\bibinfo {volume} {51}\ (\bibinfo  {publisher}
  {Springer-Verlag},\ \bibinfo {year} {2009})\BibitemShut {NoStop}%
\bibitem [{\citenamefont {Endlich}\ \emph {et~al.}(2013)\citenamefont
  {Endlich}, \citenamefont {Nicolis}, \citenamefont {Porto},\ and\
  \citenamefont {Wang}}]{Endlich:2012vt}%
  \BibitemOpen
  \bibfield  {author} {\bibinfo {author} {\bibfnamefont {S.}~\bibnamefont
  {Endlich}}, \bibinfo {author} {\bibfnamefont {A.}~\bibnamefont {Nicolis}},
  \bibinfo {author} {\bibfnamefont {R.~A.}\ \bibnamefont {Porto}}, \ and\
  \bibinfo {author} {\bibfnamefont {J.}~\bibnamefont {Wang}},\ }\href {\doibase
  10.1103/PhysRevD.88.105001} {\bibfield  {journal} {\bibinfo  {journal} {Phys.
  Rev.}\ }\textbf {\bibinfo {volume} {D88}},\ \bibinfo {pages} {105001}
  (\bibinfo {year} {2013})},\ \Eprint {http://arxiv.org/abs/1211.6461}
  {arXiv:1211.6461 [hep-th]} \BibitemShut {NoStop}%
\bibitem [{\citenamefont {Herglotz}(1911)}]{herglotz}%
  \BibitemOpen
  \bibfield  {author} {\bibinfo {author} {\bibfnamefont {G.}~\bibnamefont
  {Herglotz}},\ }\href@noop {} {\bibfield  {journal} {\bibinfo  {journal} {Ann.
  Phy.}\ }\textbf {\bibinfo {volume} {36}},\ \bibinfo {pages} {493} (\bibinfo
  {year} {1911})}\BibitemShut {NoStop}%
\bibitem [{\citenamefont {Ballesteros}\ and\ \citenamefont
  {Bellazzini}(2013)}]{Ballesteros:2012kv}%
  \BibitemOpen
  \bibfield  {author} {\bibinfo {author} {\bibfnamefont {G.}~\bibnamefont
  {Ballesteros}}\ and\ \bibinfo {author} {\bibfnamefont {B.}~\bibnamefont
  {Bellazzini}},\ }\href {\doibase 10.1088/1475-7516/2013/04/001} {\bibfield
  {journal} {\bibinfo  {journal} {JCAP}\ }\textbf {\bibinfo {volume} {1304}},\
  \bibinfo {pages} {001} (\bibinfo {year} {2013})},\ \Eprint
  {http://arxiv.org/abs/1210.1561} {arXiv:1210.1561 [hep-th]} \BibitemShut
  {NoStop}%
\bibitem [{\citenamefont {Soper}(2008)}]{soper1975}%
  \BibitemOpen
  \bibfield  {author} {\bibinfo {author} {\bibfnamefont {D.}~\bibnamefont
  {Soper}},\ }\href@noop {} {\emph {\bibinfo {title} {{Classical Field
  Theory}}}}\ (\bibinfo  {publisher} {Dover},\ \bibinfo {year}
  {2008})\BibitemShut {NoStop}%
\bibitem [{\citenamefont {Coleman}(1973)}]{Coleman:1973ci}%
  \BibitemOpen
  \bibfield  {author} {\bibinfo {author} {\bibfnamefont {S.~R.}\ \bibnamefont
  {Coleman}},\ }\href {\doibase 10.1007/BF01646487} {\bibfield  {journal}
  {\bibinfo  {journal} {Commun. Math. Phys.}\ }\textbf {\bibinfo {volume}
  {31}},\ \bibinfo {pages} {259} (\bibinfo {year} {1973})}\BibitemShut
  {NoStop}%
\bibitem [{\citenamefont {Jevicki}(1977)}]{Jevicki:1977zn}%
  \BibitemOpen
  \bibfield  {author} {\bibinfo {author} {\bibfnamefont {A.}~\bibnamefont
  {Jevicki}},\ }\href {\doibase 10.1016/0370-2693(77)90229-5} {\bibfield
  {journal} {\bibinfo  {journal} {Phys. Lett.}\ }\textbf {\bibinfo {volume}
  {B71}},\ \bibinfo {pages} {327} (\bibinfo {year} {1977})}\BibitemShut
  {NoStop}%
\bibitem [{\citenamefont {Elitzur}(1983)}]{Elitzur:1978ww}%
  \BibitemOpen
  \bibfield  {author} {\bibinfo {author} {\bibfnamefont {S.}~\bibnamefont
  {Elitzur}},\ }\href {\doibase 10.1016/0550-3213(83)90682-X} {\bibfield
  {journal} {\bibinfo  {journal} {Nucl. Phys.}\ }\textbf {\bibinfo {volume}
  {B212}},\ \bibinfo {pages} {501} (\bibinfo {year} {1983})}\BibitemShut
  {NoStop}%
\bibitem [{\citenamefont {McKane}\ and\ \citenamefont
  {Stone}(1980)}]{McKane:1979cm}%
  \BibitemOpen
  \bibfield  {author} {\bibinfo {author} {\bibfnamefont {A.}~\bibnamefont
  {McKane}}\ and\ \bibinfo {author} {\bibfnamefont {M.}~\bibnamefont {Stone}},\
  }\href {\doibase 10.1016/0550-3213(80)90396-X} {\bibfield  {journal}
  {\bibinfo  {journal} {Nucl. Phys.}\ }\textbf {\bibinfo {volume} {B163}},\
  \bibinfo {pages} {169} (\bibinfo {year} {1980})}\BibitemShut {NoStop}%
\bibitem [{\citenamefont {David}(1980)}]{David:1980gi}%
  \BibitemOpen
  \bibfield  {author} {\bibinfo {author} {\bibfnamefont {F.}~\bibnamefont
  {David}},\ }\href {\doibase 10.1016/0370-2693(80)90790-X} {\bibfield
  {journal} {\bibinfo  {journal} {Phys. Lett.}\ }\textbf {\bibinfo {volume}
  {B96}},\ \bibinfo {pages} {371} (\bibinfo {year} {1980})}\BibitemShut
  {NoStop}%
\bibitem [{\citenamefont {David}(1981)}]{David:1980rr}%
  \BibitemOpen
  \bibfield  {author} {\bibinfo {author} {\bibfnamefont {F.}~\bibnamefont
  {David}},\ }\href {\doibase 10.1007/BF01208892} {\bibfield  {journal}
  {\bibinfo  {journal} {Commun. Math. Phys.}\ }\textbf {\bibinfo {volume}
  {81}},\ \bibinfo {pages} {149} (\bibinfo {year} {1981})}\BibitemShut
  {NoStop}%
\bibitem [{Note5()}]{Note5}%
  \BibitemOpen
  \bibinfo {note} {These quantities are not all Poincar\'{e} invariant, so they
  are still really only proxies for observables.}\BibitemShut {Stop}%
\bibitem [{\citenamefont {Anastasiou}\ and\ \citenamefont
  {Lazopoulos}(2004)}]{air}%
  \BibitemOpen
  \bibfield  {author} {\bibinfo {author} {\bibfnamefont {C.}~\bibnamefont
  {Anastasiou}}\ and\ \bibinfo {author} {\bibfnamefont {A.}~\bibnamefont
  {Lazopoulos}},\ }\href {\doibase 10.1088/1126-6708/2004/07/046} {\bibfield
  {journal} {\bibinfo  {journal} {JHEP}\ }\textbf {\bibinfo {volume} {0407}},\
  \bibinfo {pages} {046} (\bibinfo {year} {2004})},\ \Eprint
  {http://arxiv.org/abs/hep-ph/0404258} {arXiv:hep-ph/0404258 [hep-ph]}
  \BibitemShut {NoStop}%
\bibitem [{\citenamefont {{Vermaseren}}()}]{form}%
  \BibitemOpen
  \bibfield  {author} {\bibinfo {author} {\bibfnamefont {J.~A.~M.}\
  \bibnamefont {{Vermaseren}}},\ }\href@noop {} {\ }\Eprint
  {http://arxiv.org/abs/math-ph/0010025} {math-ph/0010025} \BibitemShut
  {NoStop}%
\bibitem [{\citenamefont {Dubovsky}\ \emph {et~al.}(2006)\citenamefont
  {Dubovsky}, \citenamefont {Gregoire}, \citenamefont {Nicolis},\ and\
  \citenamefont {Rattazzi}}]{Dubovsky:2005xd}%
  \BibitemOpen
  \bibfield  {author} {\bibinfo {author} {\bibfnamefont {S.}~\bibnamefont
  {Dubovsky}}, \bibinfo {author} {\bibfnamefont {T.}~\bibnamefont {Gregoire}},
  \bibinfo {author} {\bibfnamefont {A.}~\bibnamefont {Nicolis}}, \ and\
  \bibinfo {author} {\bibfnamefont {R.}~\bibnamefont {Rattazzi}},\ }\href
  {\doibase 10.1088/1126-6708/2006/03/025} {\bibfield  {journal} {\bibinfo
  {journal} {JHEP}\ }\textbf {\bibinfo {volume} {0603}},\ \bibinfo {pages}
  {025} (\bibinfo {year} {2006})},\ \Eprint
  {http://arxiv.org/abs/hep-th/0512260} {arXiv:hep-th/0512260 [hep-th]}
  \BibitemShut {NoStop}%
\bibitem [{\citenamefont {Thomson}(1880)}]{Thomson}%
  \BibitemOpen
  \bibfield  {author} {\bibinfo {author} {\bibfnamefont {W.}~\bibnamefont
  {Thomson}},\ }\href@noop {} {\bibfield  {journal} {\bibinfo  {journal} {Phil.
  Mag.}\ }\textbf {\bibinfo {volume} {10}},\ \bibinfo {pages} {155} (\bibinfo
  {year} {1880})}\BibitemShut {NoStop}%
\bibitem [{\citenamefont {Endlich}\ and\ \citenamefont
  {Nicolis}(2013)}]{Endlich:2013dma}%
  \BibitemOpen
  \bibfield  {author} {\bibinfo {author} {\bibfnamefont {S.}~\bibnamefont
  {Endlich}}\ and\ \bibinfo {author} {\bibfnamefont {A.}~\bibnamefont
  {Nicolis}},\ }\href@noop {} {\  (\bibinfo {year} {2013})},\ \Eprint
  {http://arxiv.org/abs/1303.3289} {arXiv:1303.3289 [hep-th]} \BibitemShut
  {NoStop}%
\bibitem [{\citenamefont {{Mart{\'{\i}}n-Garc{\'{\i}}a}}\ \emph
  {et~al.}(2007)\citenamefont {{Mart{\'{\i}}n-Garc{\'{\i}}a}}, \citenamefont
  {{Portugal}},\ and\ \citenamefont {{Manssur}}}]{xact}%
  \BibitemOpen
  \bibfield  {author} {\bibinfo {author} {\bibfnamefont {J.~M.}\ \bibnamefont
  {{Mart{\'{\i}}n-Garc{\'{\i}}a}}}, \bibinfo {author} {\bibfnamefont
  {R.}~\bibnamefont {{Portugal}}}, \ and\ \bibinfo {author} {\bibfnamefont
  {L.~R.~U.}\ \bibnamefont {{Manssur}}},\ }\href {\doibase
  10.1016/j.cpc.2007.05.015} {\bibfield  {journal} {\bibinfo  {journal} {Comp.
  Phys. Commun.}\ }\textbf {\bibinfo {volume} {177}},\ \bibinfo {pages} {640}
  (\bibinfo {year} {2007})},\ \Eprint {http://arxiv.org/abs/0704.1756}
  {arXiv:0704.1756} \BibitemShut {NoStop}%
\end{thebibliography}%

\end{document}